\documentclass[aps,prb,twocolumn,showpacs]{revtex4}

\usepackage{graphicx}
\usepackage{amsmath}
\usepackage{amssymb}

\begin{document}

\title{Dynamical spin response in cuprate superconductors from low-energy to high-energy}

\author{L\"ulin Kuang}

\affiliation{Department of Physics, Beijing Normal University, Beijing 100875, China}

\author{Yu Lan}

\affiliation{Department of Physics and Electronic Information Science, Hengyang Normal University, Hengyang 421002, China}

\author{Shiping Feng}
\email{spfeng@bnu.edu.cn}

\affiliation{Department of Physics, Beijing Normal University, Beijing 100875, China~~~~}

\begin{abstract}
Within the framework of the kinetic energy driven superconducting mechanism, the dynamical spin response of cuprate superconductors is studied from low-energy to high-energy. The spin self-energy is evaluated explicitly in terms of the collective charge carrier modes in the particle-hole and particle-particle channels, and employed to calculate the dynamical spin structure factor. Our results show the existence of damped but well-defined dispersive spin excitations in the whole doping phase diagram. In particular, the low-energy spin excitations in the superconducting-state have an hour-glass-shaped dispersion, with commensurate resonance that appears in the superconducting-state {\it only}, while the low-energy incommensurate spin fluctuations can persist into the normal-state. The high-energy spin excitations in the superconducting-state on the other hand retain roughly constant energy as a function of doping, with spectral weights and dispersion relations comparable to those in the corresponding normal-state. The theory also shows that the unusual magnetic correlations in cuprate superconductors can be ascribed purely to the spin self-energy effects which arise directly from the charge carrier-spin interaction in the kinetic energy of the system.
\end{abstract}


\maketitle

\section{Introduction}\label{Introduction}

The interplay between antiferromagnetism and superconductivity is one of the challenging issues in cuprate superconductors \cite{Anderson87,Kastner98,Fujita12}. This follows from a fact that the parent compounds of cuprate superconductors are a form of non-conductor called a Mott insulator with an antiferromagnetic (AF) long-range order (AFLRO) \cite{Kastner98,Fujita12}, where a single common structural feature is the presence of the CuO$_{2}$ planes. Furthermore, the inelastic neutron scattering (INS) experiments have shown that the low-energy spin excitations in these parent compounds are well described by an AF Heisenberg model \cite{Kastner98,Fujita12,Hayden91,Hayden96,Coldea01} with magnetic exchange coupling constant $J\sim 0.1$ eV. The spin excitations with AFLRO are called as magnons. When the CuO$_{2}$ planes are doped with charge carriers, the AFLRO phase subsides and superconductivity emerges leaving the AF short-range order (AFSRO) correlations still intact. This AFSRO can still support spin waves, but the spin excitations with AFSRO are damped. The damped spin excitations are known as paramagnons \cite{Miller11}. In particular, in the doped regime, the charge carriers couple to the spin excitations \cite{Eschrig06,Fong95}, and it has been argued that if the spin excitations exist over a wide enough range of energies, they can produce the necessary attractive interaction to induce superconductivity \cite{Miller11,Eschrig06,Fong95}.

The early INS measurements \cite{Fujita12,Eschrig06,Fong95,Birgeneau89,Cheong91,Yamada98,Dai01,Wakimoto04,Hayden04,Tranquada04} on cuprate superconductors have demonstrated that the doped charge carriers cause substantial changes to the low-energy spin excitation spectrum, and a consistent pattern has been identified as the {\it hour-glass-shaped} dispersion.  This hour-glass-shaped dispersion was first observed in the spin excitations of YBa$_{2}$Cu$_{3}$O$_{6.6}$ \cite{Hayden04} and La$_{1.875}$Ba$_{0.125}$CuO$_{4}$ \cite{Tranquada04}, where two incommensurate (IC) components of the low-energy spin excitation spectrum are separated by a commensurate resonance energy $\omega_{\rm r}$ at the waist of the hour glass. In the upward component, above the commensurate resonance energy $\omega_{\rm r}$, the spin excitation spectrum is similar to what one would expect from AF spin fluctuations with a finite gap, and it is relevant to the results for different families of cuprate superconductors that appear to scale with the magnetic exchange coupling constant $J$ for the parent Mott insulators. Moreover, for a given excitation energy, the magnetic scattering peaks lie on a circle of radius of $\bar{\delta}_{\rm IC}'$, with the incommensurability parameter $\bar{\delta}_{\rm IC}'$ that is defined as a deviation of the peak position from the AF wave vector $[1/2,1/2]$ (hereafter we use the units of $[2\pi,2\pi]$) in the Brillouin zone (BZ), and then the distribution of the spectral weight of the IC magnetic scattering peaks is rather isotropic. On the other hand, in the downward component, below $\omega_{\rm r}$, the distribution of the spectral weight of the IC magnetic scattering peaks is quite anisotropic \cite{Fujita12,Eschrig06,Fong95,Birgeneau89,Cheong91,Yamada98,Dai01,Wakimoto04,Hayden04,Tranquada04}. In particular, it is remarkable \cite{Fujita12,Eschrig06,Fong95,Birgeneau89,Cheong91,Yamada98,Dai01,Wakimoto04,Hayden04,Tranquada04,Hinkov04,Bourges05,He01,Bourges00,Arai99} that in analogy to the domelike shape of the doping dependence of the SC transition temperature $T_{\rm c}$, the commensurate resonance energy $\omega_{\rm r}$ increases with increasing doping in the underdoped regime, and reaches a maximum around the optimal doping, then decreases in the overdoped regime, reflecting a intrinsical relationship between $\omega_{\rm r}$ and $T_{\rm c}$. Although the IC magnetic scattering has been also observed in the normal-state, the commensurate resonance is a new feature that appears in the SC-state {\it only} \cite{Fujita12,Eschrig06,Fong95,Birgeneau89,Cheong91,Yamada98,Dai01,Wakimoto04,Hayden04,Tranquada04,Hinkov04,Bourges05,He01,Bourges00,Arai99}. Later, this hour-glass-shaped dispersion was found in several different families of cuprate superconductors \cite{Vignolle07,Stock05,Stock10,Xu09}. However, because of technical limitations, only the low-energy ($E\sim 10-80$ meV) spin excitations in a small range of momentum space around the AF wave vector are detected by the INS measurements \cite{Eschrig06,Fong95,Birgeneau89,Cheong91,Yamada98,Dai01,Wakimoto04,Hayden04,Tranquada04,Hinkov04,Bourges05,He01,Bourges00,Arai99,Vignolle07,Stock05,Stock10,Xu09}, and they may be insufficient to produce superconductivity \cite{Miller11,Eschrig06,Vojta11}. In recent years, instrumentation for resonant inelastic X-ray scattering (RIXS) with both soft and hard X-rays has improved dramatically, allowing this technique to directly measure the high-energy ($E\sim 80-500$ meV) spin excitations of cuprate superconductors in the wide energy-momentum window that cannot be detected by the INS measurements \cite{Dean14}. In this case, as a compensation for the miss of a significant part of the spectral weight of the spin excitations in the INS studies \cite{Eschrig06,Fong95,Birgeneau89,Cheong91,Yamada98,Dai01,Wakimoto04,Hayden04,Tranquada04,Hinkov04,Bourges05,He01,Bourges00,Arai99,Vignolle07,Stock05,Stock10,Xu09}, the RIXS experiments \cite{Dean14,Tacon11,Dean13,Dean13a,Tacon13} have purported to measure high-energy spin excitations in a large family of cuprate superconductors that persist well into the overdoped regime and bear a striking resemblance to those found in the parent compound \cite{Braicovich10,Guarise10,Piazza12}, indicating that a local-moment picture accounts for the observed spin excitations at elevated energies even up to the overdoped regime. In particular, the very importance is that the combined these RIXS-INS experimental data have identified the spin excitations with high intensity over a large part of moment space, and shown that the spin excitations exist across the entire range of the SC dome, and with sufficient intensity to mediate superconductivity in cuprate superconductors \cite{Miller11,Eschrig06,Vojta11}.

Although the spin excitation spectrum of cuprate superconductors from low-energy to high-energy in the whole doping phase diagram is well-established from the INS \cite{Fujita12,Eschrig06,Fong95,Birgeneau89,Cheong91,Yamada98,Dai01,Wakimoto04,Hayden04,Tranquada04,Hinkov04,Bourges05,He01,Bourges00,Arai99,Vignolle07,Stock05,Stock10,Xu09} and RIXS \cite{Dean14,Tacon11,Dean13,Dean13a,Tacon13} measurements, its full understanding is still a challenging issue. In our early studies \cite{Feng98,Feng04}, the low-energy spin excitations of the underdoped cuprate superconductors in the normal-state has been discussed by considering spin fluctuations around the mean-field (MF) solution, where the spin self-energy is evaluated from the charge carrier bubble in the particle-hole channel, and then the obtained results are qualitatively consistent with the corresponding experimental data. In this paper, as a complement of our previous analysis of the low-energy spin excitations of the underdoped cuprate superconductors in the normal-state, we start from the kinetic energy driven SC mechanism \cite{Feng0306,Guo06,Feng08} to discuss the dynamical spin response of cuprate superconductors from low-energy to high-energy in both the SC- and normal-states, where one of our main results is that the both damped but well-defined dispersive low-energy and high-energy spin excitations exist across the whole doping phase diagram. The low-energy spin excitations are strongly renormalized due to the charge carrier-spin interaction to form an hour-glass-shaped dispersion in the SC-state. In particular, we identify that the commensurate resonance is closely related to the process of the creation of charge carrier pairs, and appears in the SC-state only, while the low-energy IC magnetic scattering is mainly associated with the motion of charge carrier quasiparticles, and therefore can persist into the normal-state. On the other hand, the charge carrier doping has a more modest effect on the high-energy spin excitations, and then the high-energy spin fluctuations bear a striking resemblance to those found in the parent compound.

The paper is organized as follows. The basic formalism is presented in Sec. \ref{framework}, where we evaluate explicitly the full spin Green's function (then the spin self-energy) in the SC-state in terms of the collective charge carrier modes in the particle-hole and particle-particle channels. The dynamical spin structure factor, however, is obtained from the imaginary part of the full spin Green's function, and is employed to discuss the quantitative characteristics of the dynamical spin response in cuprate superconductors in Sec. \ref{sc-state} for the SC-state and Sec. \ref{normal-state} for the normal-state, where we show that although the magnetic scattering of spins dominates the spin dynamics, the effect of charge carriers on the spin part in terms of the spin self-energy renormalization is critical in determining the doping and energy dependence of the dynamical spin response in cuprate superconductors.

\section{Dynamical spin response in cuprate superconductors}\label{framework}

Since the single common element in cuprate superconductors is the presence of the CuO$_{2}$ planes as mentioned above, it is believed that the anomalous properties are closely related to the doped CuO$_{2}$ planes \cite{Anderson87,Dagotto94,Phillips10}. In particular, as originally emphasized by Anderson \cite{Anderson87}, the essential physics of the doped CuO$_{2}$ plane is properly captured by the $t$-$J$ model on a square lattice,
\begin{eqnarray}\label{tjham}
H&=&-t\sum_{l\hat{\eta}\sigma}C^{\dagger}_{l\sigma}C_{l+\hat{\eta}\sigma}+t'\sum_{l\hat{\tau}\sigma}C^{\dagger}_{l\sigma}C_{l+\hat{\tau}\sigma}\nonumber\\
&+&\mu\sum_{l\sigma}C^{\dagger}_{l\sigma} C_{l\sigma}+J\sum_{l\hat{\eta}}{\bf S}_{l}\cdot {\bf S}_{l+\hat{\eta}},
\end{eqnarray}
where the summation is over all sites $l$, and for each $l$, over its nearest-neighbors $\hat{\eta}$ or the next nearest-neighbors $\hat{\tau}$, $C^{\dagger}_{l\sigma}$ ($C_{l\sigma}$) is the electron creation (annihilation) operator, ${\bf S}_{l}=(S^{\rm x}_{l},S^{\rm y}_{l}, S^{\rm z}_{l})$ are spin operators, $t$ ($t'$) is the nearest-neighbor (next nearest-neighbor) hopping integral, $J$ is the AF exchange coupling constant, and $\mu$ is the chemical potential. In this $t$-$J$ model (\ref{tjham}), the kinetic energy term describes the motion of charge carriers, while the Heisenberg term describes the AF coupling between localized spins. In particular, the nearest-neighbor hopping integral $t$ in the kinetic energy term is much larger than the AF exchange coupling constant $J$ in the Heisenberg term, and then the AF spin configuration is strongly affected by the motion of charge carriers. The high complexity in the $t$-$J$ model comes mainly from the local constraint of no double electron occupancy, i.e., $\sum_{\sigma}C^{\dagger}_{l\sigma} C_{l\sigma}\leq 1$, which can be treated properly within the charge-spin separation (CSS) fermion-spin theory \cite{Feng04,Feng94,Feng08}, where the constrained electron operators $C_{l\uparrow}$ and $C_{l\downarrow}$ are decoupled as,
\begin{eqnarray}\label{css}
C_{l\uparrow}=h^{\dagger}_{l\uparrow}S^{-}_{l},~~~~C_{l\downarrow}=h^{\dagger}_{l\downarrow}S^{+}_{l},
\end{eqnarray}
respectively, with the fermion operator $h_{l\sigma}=e^{-i\Phi_{l\sigma}}h_{l}$ that keeps track of the charge degree of freedom together with some effects of spin configuration rearrangements due to the presence of the doped hole itself (charge carrier), while the spin operator $S_{l}$ represents the spin degree of freedom, and then the local constraint of no double electron occupancy is always satisfied in analytical calculations. In this CSS fermion-spin representation (\ref{css}), the original $t$-$J$ model (\ref{tjham}) can be expressed explicitly as,
\begin{eqnarray}\label{cssham}
H&=&t\sum_{l\hat{\eta}}(h^{\dagger}_{l+\hat{\eta}\uparrow}h_{l\uparrow}S^{+}_{l}S^{-}_{l+\hat{\eta}}+h^{\dagger}_{l+\hat{\eta}\downarrow}h_{l\downarrow}S^{-}_{l}
S^{+}_{l+\hat{\eta}})\nonumber\\
&-&t'\sum_{l\hat{\tau}}(h^{\dagger}_{l+\hat{\tau}\uparrow}h_{l\uparrow}S^{+}_{l}S^{-}_{l+\hat{\tau}}+h^{\dagger}_{l+\hat{\tau}\downarrow}h_{l\downarrow}
S^{-}_{l}S^{+}_{l+\hat{\tau}})
\nonumber\\
&-&\mu\sum_{l\sigma} h^{\dagger}_{l\sigma}h_{l\sigma}+J_{{\rm eff}}\sum_{l\hat{\eta}}{\bf S}_{l}\cdot {\bf S}_{l+\hat{\eta}},
\end{eqnarray}
where $S^{-}_{l}=S^{\rm x}_{l}-iS^{\rm y}_{l}$ and $S^{+}_{l}=S^{\rm x}_{l}+iS^{\rm y}_{l}$ are the spin-lowering and spin-raising operators for the spin $S=1/2$, respectively, $J_{{\rm eff}}=(1-\delta)^{2}J$, and $\delta=\langle h^{\dagger}_{l\sigma}h_{l\sigma}\rangle=\langle h^{\dagger}_{l}h_{l}\rangle$ is the charge carrier doping concentration. As a consequence, the kinetic energy in the $t$-$J$ model (\ref{cssham}) has been released as the charge carrier-spin interaction, which reflects that even the kinetic energy in the $t$-$J$ model (\ref{tjham}) has strong Coulombic contribution due to the restriction of no double electron occupancy of a given site, and therefore dominates the essential physics of cuprate superconductors.

Superconductivity, the dissipationless flow of electrical current, is a striking manifestation of a subtle form of quantum rigidity on the  macroscopic scale. It is commonly believed that the electron Cooper pairs are crucial because these electron Cooper pairs behave as effective bosons, and can form something analogous to a Bose condensate that flows without resistance \cite{Anderson07}. This follows from a fact that although electrons repel each other because of the Coulomb interaction, at low energies there can be an effective attraction that originates by the exchange of bosons \cite{Monthoux07}. In conventional superconductors, as explained by the Bardeen-Cooper-Schrieffer (BCS) theory \cite{Bardeen57}, these exchanged bosons are {\it phonons} that act like a bosonic {\it glue} to hold the electron pairs together \cite{Cooper56}, then these electron Cooper pairs condense into a coherent macroscopic quantum state that is insensitive to impurities and imperfections and hence conducts electricity without resistance. As in the case of conventional superconductors, the pairing of electrons in cuprate superconductors occurs at $T_{\rm c}$, creating an energy gap in the electron excitation spectrum that serves as the SC order parameter \cite{Tsuei00}. The BCS theory is not specific to a phonon-mediated interaction, other excitations can also serve as the pairing glue \cite{Miller11,Monthoux07}. As we have mentioned in Sec. \ref{Introduction}, the experimental results from the INS \cite{Eschrig06,Fong95,Birgeneau89,Cheong91,Yamada98,Dai01,Wakimoto04,Hayden04,Tranquada04,Hinkov04,Bourges05,He01,Bourges00,Arai99,Vignolle07,Stock05,Stock10,Xu09} and RIXS \cite{Dean14,Tacon11,Dean13,Dean13a,Tacon13} measurements have provided a clear link between the electron Cooper pairing mechanism and spin excitations, then a question is raised whether the spin excitations, which is a generic consequence of the strong Coulomb interaction, can mediate electron Cooper pairing in cuprate superconductors in analogy to the phonon-mediate pairing mechanism in conventional superconductors \cite{Miller11,Monthoux07}? Within the $t$-$J$ model (\ref{cssham}), we have developed a kinetic energy driven SC mechanism \cite{Feng0306}, where cuprate superconductors involve charge carrier pairs bound together by the exchange of spin excitations, then the electron Cooper pairs originating from charge carrier pairs are due to charge-spin recombination, and they condense to the d-wave SC ground-state. In particular, one of the striking features in this kinetic energy driven SC mechanism \cite{Feng0306} is that the AFSRO correlations coexist with superconductivity. The physical picture of the kinetic energy driven SC mechanism is also quite clear \cite{Feng0306}: at half-filling, each lattice site is singly occupied by a spin-up or spin-down electron, then the electron spins are coupled antiferromagnetically with AFLRO. With doping, in particular, in the doped regime without an AFLRO, the charge carriers move in the spin liquid background, and form pairs at low temperatures in the particle-particle channel induced by the charge carrier-spin interaction directly from the kinetic energy of the $t$-$J$ model (\ref{cssham}) by exchanging spin excitations in higher powers of the doping concentration, then these charge carrier pairs (then the electron Cooper pairs) condense to the SC-state. Our following work builds on the kinetic energy driven SC mechanism in Ref. \onlinecite{Feng0306}, and only a short summary of the formalism is therefore given. In our previous discussions, the full charge carrier diagonal and off-diagonal Green's functions of the $t$-$J$ model (\ref{cssham}) in the SC-state have been given explicitly as \cite{Feng0306,Guo06,Feng08},
\begin{subequations}\label{BCSHGF}
\begin{eqnarray}
g({\bf k},\omega)&=&Z_{\rm hF}\left ({U^{2}_{{\rm h}{\bf k}}\over\omega-E_{{\rm h}{\bf k}}}+{V^{2}_{{\rm h}{\bf k}}\over\omega+E_{{\rm h}{\bf k}}}\right ),\label{BCSHDGF}\\
\Gamma^{\dagger}({\bf k},\omega)&=&-Z_{\rm hF}{\bar{\Delta}_{\rm hZ}({\bf k})\over 2E_{{\rm h}{\bf k}}}\left ({1\over \omega-E_{{\rm h}{\bf k}}}-{1\over\omega
+E_{{\rm h}{\bf k}}}\right ),~~~~~~~\label{BCSHODGF}
\end{eqnarray}
\end{subequations}
where the charge carrier quasiparticle spectrum $E_{{\rm h}{\bf k}}=\sqrt{\bar{\xi}^{2}_{{\bf k}}+\mid\bar{\Delta}_{\rm hZ}({\bf k})\mid^{2}}$, the charge carrier quasiparticle coherence factors $U^{2}_{{\rm h}{\bf k}}=[1+{\bar{\xi_{{\bf k}}}/E_{{\rm h}{\bf k}}}]/2$ and $V^{2}_{{\rm h} {\bf k}}=[1-{\bar{\xi_{{\bf k}}}/E_{{\rm h}{\bf k}}}]/2$,
the renormalized charge carrier excitation spectrum $\bar{\xi}_{{\bf k}}=Z_{\rm hF}\xi_{{\bf k}}$, the charge carrier excitation spectrum $\xi_{\bf k}=Zt\chi_{1}\gamma_{{\bf k}} -Zt'\chi_{2}\gamma_{{\bf k}}'-\mu$, with the spin correlation functions $\chi_{1}=\langle S^{+}_{l}S^{-}_{l+\hat{\eta}}\rangle$ and $\chi_{2}=\langle S_{l}^{+}S_{l+\hat{\tau}}^{-} \rangle$, $\gamma_{{\bf k}}=(1/Z)\sum_{\hat{\eta}}e^{i{\bf k}\cdot\hat{\eta}}$, $\gamma_{{\bf k}}'=(1/Z)\sum_{\hat{\tau}}e^{i{\bf k}\cdot\hat{\tau}}$, and $Z$ is the number of the nearest-neighbor or next nearest-neighbor sites on a square lattice, the renormalized charge carrier pair gap $\bar{\Delta}_{\rm hZ}({\bf k})=Z_{\rm hF} \bar{\Delta}_{\rm h}({\bf k})$, while the charge carrier pair gap is closely related to the charge carrier self-energy in the particle-particle channel as $\bar{\Delta}_{\rm h}({\bf k})=\Sigma^{({\rm h})}_{2}({\bf k},\omega=0)$, and has a d-wave form $\bar{\Delta}_{\rm h}({\bf k})=\bar{\Delta}_{\rm h}\gamma^{(\rm d)}_{{\bf k}}$ with $\gamma^{(\rm d)}_{{\bf k}}= ({\rm cos}k_{x}-{\rm cos}k_{y})/2$. The charge carrier quasiparticle coherent weight is obtained from the antisymmetric part of the charge carrier self-energy in the particle-hole channel as $Z^{-1}_{\rm hF}=1-{\rm Re}\Sigma^{({\rm h}) }_{\rm 1o}({\bf k}_{0},\omega=0)$ with the wave vector ${\bf k}_{0}$ that has been chosen as ${\bf k}_{0}=[0.5,0]$ just as it has been done in the experiments \cite{Ding01}. The charge carrier self-energies $\Sigma^{({\rm h})}_{1}({\bf k},\omega)$ in the particle-hole channel and $\Sigma^{({\rm h})}_{2}({\bf k},\omega)$ in the particle-particle channel have been evaluated from the spin bubble as \cite{Feng0306,Guo06,Feng08},
\begin{widetext}  
\begin{subequations}\label{SE}
\begin{eqnarray}
\Sigma^{({\rm }h)}_{1}({\bf k},i\omega_{n})&=&{1\over N^{2}}\sum_{{\bf p,p'}}\Lambda^{2}_{{\bf p}+{\bf p}'+{\bf k}}{1\over \beta}\sum_{ip_{m}}g({\bf p}+{\bf k},ip_{m}+i\omega_{n}) \Pi({\bf p},{\bf p}',ip_{m}), \label{self-energy-1}\\
\Sigma^{({\rm h})}_{2}({\bf k},i\omega_{n})&=&{1\over N^{2}}\sum_{{\bf p,p'}}\Lambda^{2}_{{\bf p}+{\bf p}'+{\bf k}}{1\over \beta}\sum_{ip_{m}}\Gamma^{\dagger}({\bf p}+{\bf k}, ip_{m}+i\omega_{n})\Pi({\bf p},{\bf p}',ip_{m}), \label{self-energy-2}
\end{eqnarray}
\end{subequations}
\end{widetext} 
respectively, with $\Lambda_{{\bf k}}=Zt\gamma_{\bf k}-Zt'\gamma_{\bf k}'$, and the spin bubble,
\begin{eqnarray}\label{SB}
\Pi({\bf p},{\bf p}',ip_{m})&=&{1\over\beta}\sum_{ip'_{m}}D^{(0)}({\bf p'},ip_{m}')\nonumber\\
&\times&D^{(0)}({\bf p}'+{\bf p},ip_{m}'+ip_{m}),
\end{eqnarray}
where the MF spin Green's function has been obtained explicitly as \cite{Feng0306,Guo06,Feng08},
\begin{eqnarray}
D^{(0)}({\bf k},\omega)&=&{B_{\bf k}\over 2\omega_{\bf k}}\left ({1\over \omega-\omega_{\bf k}}-{1\over\omega+\omega_{\bf k}}\right ),\label{MFSGF}
\end{eqnarray}
with the MF spin excitation spectrum $\omega_{\bf k}$, and the function $B_{\bf k}$ have been given in Refs. \cite{Guo06,Feng08}. In particular, it should be emphasized that all the order parameters and chemical potential $\mu$ in the above calculation have been determined by self-consistent calculation without using any adjustable parameters \cite{Feng0306,Guo06,Feng08}.

In the framework of the CSS fermion-spin theory (\ref{css}), the charge transport is mainly governed by the scattering of charge carriers due to the spin fluctuations \cite{Feng97}, while the scattering of spins due to the charge carrier fluctuations dominates the spin dynamics \cite{Feng98,Feng04}. For the discussion of the dynamical spin response in cuprate superconductors, we need to calculate the full spin Green's function, which can be expressed as,
\begin{eqnarray}\label{FSGF}
D({\bf k},\omega)={1\over D^{(0)-1}({\bf k},\omega)-\Sigma^{({\rm s})}({\bf k},\omega)}.
\end{eqnarray}
In the SC-state, the spin fluctuations occur in the charge carrier quasiparticle background, and then the spin self-energy in the SC-state can be obtained within the framework of the equation of motion method in terms of the collective charge carrier modes in the particle-hole and particle-particle channels as,
\begin{widetext} 
\begin{eqnarray}\label{SSF1}
\Sigma^{({\rm s})}({\bf k},ip_{m})&=&-{1\over N^{2}}\sum_{\bf pq}(\Lambda^{2}_{{\bf k}-{\bf p}}+\Lambda^{2}_{{\bf p}+{\bf q}+{\bf k}}){1\over\beta}\sum_{iq_{m}}
D^{(0)}({\bf q}+{\bf k},iq_{m}+ip_{m})[\Pi^{(s)}_{gg}({\bf p},{\bf q},iq_{m})-\Pi^{(s)}_{\Gamma\Gamma}({\bf p},{\bf q},iq_{m})],~~~~~~~
\end{eqnarray}
\end{widetext} 
where the charge carrier bubble $\Pi^{(s)}_{gg}({\bf p},{\bf q},iq_{m})$ in the particle-hole channel is obtained from the full charge carrier diagonal Green's function (\ref{BCSHDGF}) as,
\begin{eqnarray}\label{SBPH}
\Pi^{(s)}_{gg}({\bf p},{\bf q},iq_{m})&=&{1\over\beta}\sum_{i\omega_{n}}g({\bf p},i\omega_{n})g({\bf p}+{\bf q},i\omega_{n}+iq_{m}), ~~~~~~~
\end{eqnarray}
and is closely related to the motion of charge carrier quasiparticles, while the charge carrier bubble $\Pi^{(s)}_{\Gamma\Gamma}({\bf p},{\bf q},iq_{m})$ in the particle-particle channel is obtained from the full charge carrier off-diagonal Green's function (\ref{BCSHODGF}) as,
\begin{eqnarray}\label{SBPP}
\Pi^{(s)}_{\Gamma\Gamma}({\bf p},{\bf q},iq_{m})&=&{1\over\beta}\sum_{i\omega_{n}}\Gamma^{\dagger}({\bf p},i\omega_{n})\Gamma({\bf p}+{\bf q},i\omega_{n}+iq_{m}),~~~~~~~
\end{eqnarray}
and therefore is directly associated with the creation of charge carrier pairs. Substituting the full charge carrier Green's function (\ref{BCSHGF}) and the MF spin Green's function (\ref{MFSGF}) into Eqs. (\ref{SBPH}), (\ref{SBPP}), and (\ref{SSF1}), we then evaluate the spin self-energy explicitly in the SC-state as,
\begin{widetext} 
\begin{eqnarray}\label{SSF}
\Sigma^{({\rm s})}({\bf k},\omega)&=&-{1\over 2N^{2}}\sum_{\bf pq,\nu=1,2}(-1)^{\nu+1}\Omega({\bf k},{\bf p},{\bf q})\left ({I_{+}({\bf p},{\bf q})F^{\rm (s)}_{\nu+}({\bf k},{\bf p}, {\bf q})\over\omega^{2}-[\omega_{{\bf q}+ {\bf k}}- (-1)^{\nu+1}(E_{{\rm h}{\bf p}+{\bf q}}-E_{{\rm h}{\bf p}})]^{2}}\right .\nonumber\\
&+& \left . {I_{-}({\bf p},{\bf q})F^{\rm (s)}_{\nu-}({\bf k},{\bf p},{\bf q})\over\omega^{2}-[\omega_{{\bf q}+{\bf k}}- (-1)^{\nu+1}(E_{{\rm h}{\bf p}+{\bf q}}+E_{{\rm h}{\bf p}}) ]^{2}} \right ),
\end{eqnarray}
where $\Omega({\bf k},{\bf p},{\bf q})=Z^{2}_{\rm hF}(\Lambda^{2}_{{\bf k}-{\bf p}}+\Lambda^{2}_{{\bf p}+{\bf q}+{\bf k}})B_{{\bf q}+{\bf k}}/(2\omega_{{\bf q}+{\bf k}})$, the charge carrier coherence factors for the processes,
\begin{subequations}\label{CFS}
\begin{eqnarray}
I_{+}({\bf p},{\bf q})&=&1+{\bar{\xi}_{\bf p}\bar{\xi}_{{\bf p}+{\bf q}}-\bar{\Delta}_{\rm hZ}({\bf p})\bar{\Delta}_{\rm hZ}({\bf p}+{\bf q})\over E_{{\rm h}{\bf p}}E_{{\rm h}{\bf p} +{\bf q}}},\label{CFS1}\\
I_{-}({\bf p},{\bf q})&=&1-{\bar{\xi}_{\bf p}\bar{\xi}_{{\bf p}+{\bf q}}-\bar{\Delta}_{\rm hZ}({\bf p})\bar{\Delta}_{\rm hZ}({\bf p}+{\bf q})\over E_{{\rm h}{\bf p}}E_{{\rm h}{\bf p} +{\bf q}}},\label{CFS2}
\end{eqnarray}
\end{subequations}
and the functions,
\begin{subequations}
\begin{eqnarray}
F^{\rm (s)}_{\nu+}({\bf k},{\bf p},{\bf q})&=&[\omega_{{\bf q}+{\bf k}}-(-1)^{\nu+1}(E_{{\rm h}{\bf p}+{\bf q}}-E_{{\rm h}{\bf p}})]\{n_{\rm B}(\omega_{{\bf q}+{\bf k}})[n_{\rm F} (E_{{\rm h}{\bf p}})-n_{\rm F}(E_{{\rm h}{\bf p}+{\bf q}})]\nonumber\\
&-&(-1)^{\nu+1}n_{\rm F}[(-1)^{\nu}E_{{\rm h}{\bf p}}]n_{\rm F}[(-1)^{\nu+1}E_{{\rm h}{\bf p}+{\bf q}}]\},\\
F^{\rm (s)}_{\nu-}({\bf k},{\bf p},{\bf q})&=& [\omega_{{\bf q}+{\bf k}}-(-1)^{\nu+1}(E_{{\rm h}{\bf p}+{\bf q}}+E_{{\rm h}{\bf p}})]\{n_{\rm B}(\omega_{{\bf q}+{\bf k}})[1-n_{\rm F} (E_{{\rm h}{\bf p}})-n_{\rm F}(E_{{\rm h}{\bf p}+{\bf q}})]\nonumber\\
&-&(-1)^{\nu+1}n_{\rm F}[(-1)^{\nu+1}E_{{\rm h}{\bf p}}]n_{\rm F}[(-1)^{\nu+1}E_{{\rm h}{\bf p}+{\bf q}}]\},
\end{eqnarray}
\end{subequations}
where $n_{\rm B}(\omega)$ and $n_{\rm F}(\omega)$ are the boson and fermion distribution functions, respectively.

With the help of the full spin Green's function (\ref{FSGF}), the dynamical spin structure factor of cuprate superconductors in the SC-state is obtained as  \cite{Feng98,Feng04},
\begin{eqnarray}\label{DSSF}
S({\bf k},\omega)&=&-2[1+n_{\rm B}(\omega)]{\rm Im}D({\bf k},\omega)
=-{2[1+n_{\rm B}(\omega)]B^{2}_{{\bf k}}{\rm Im}\Sigma^{({\rm s})}({\bf k},\omega)\over [\omega^{2}-\omega^{2}_{\bf k}-B_{{\bf k}}{\rm Re}\Sigma^{({\rm s})}({\bf k},\omega)]^{2} +[B_{{\bf k}}{\rm Im}\Sigma^{({\rm s})}({\bf k},\omega)]^{2}}, ~~~~~
\end{eqnarray}
where ${\rm Im}\Sigma^{({\rm s})}({\bf k},\omega)$ and ${\rm Re}\Sigma^{({\rm s})}({\bf k},\omega)$ are the corresponding imaginary and real parts of the spin self-energy (\ref{SSF}), respectively.
\end{widetext} 

\section{Quantitative characteristics in the SC-state}\label{sc-state}

In this section, we present some quantitative characteristics of the dynamical spin response of cuprate superconductors in the SC-state. In cuprate superconductors, although the values of $J$, $t$, and $t'$ are believed to vary somewhat from compound to compound, the commonly used parameters in this paper are chosen as $t/J=2.5$, $t'/t=0.3$, and $J=100$ meV as in our previous discussions \cite{Feng0306,Guo06,Feng08}.

\subsection{Universal Low-energy spin excitation spectrum}

\begin{figure*}[t!]  
\includegraphics[scale=0.9]{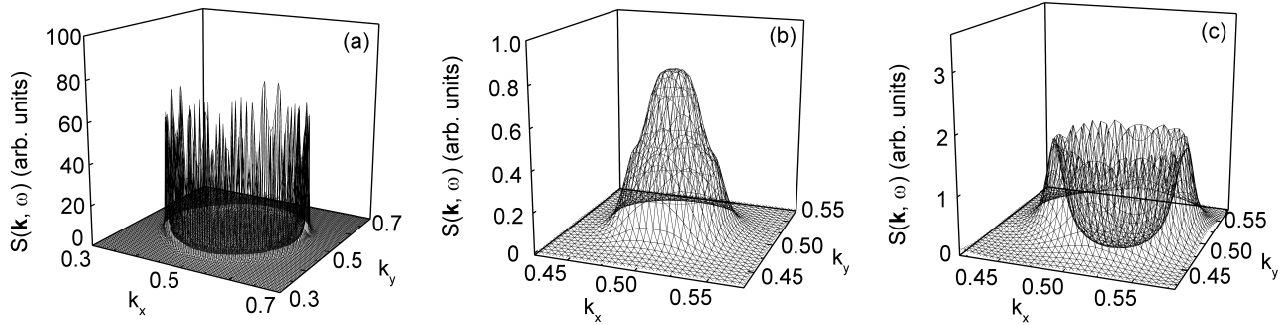}
\caption{The dynamical spin structure factor $S({\bf k},\omega)$ in the $[k_{x},k_{y}]$ plane at $\delta=0.21$ with $T=0.002J$ for $t/J=2.5$ and $t'/t=0.3$ in (a) $\omega=0.1J$, (b) $\omega=0.3J$, and (c) $\omega =0.5J$. \label{fig1}}
\end{figure*}

\begin{figure}[h!]
\includegraphics[scale=0.4]{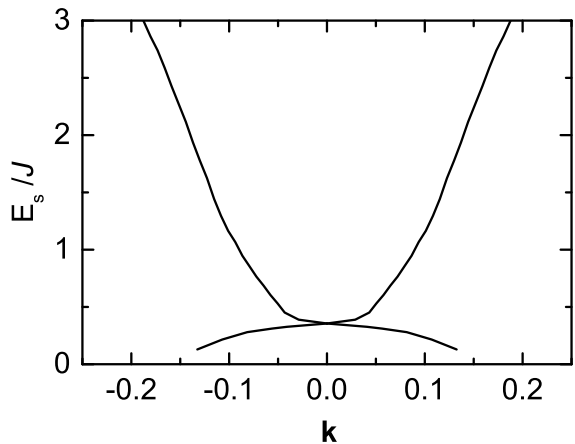}
\caption{The energy dependence of the position of the magnetic scattering peaks at $\delta=0.21$ with $T=0.002J$ for $t/J=2.5$ and $t'/t=0.3$. \label{fig2}}
\end{figure}

Firstly, we discuss unusual feature of the low-energy magnetic scattering. Of course, at half-filling, the parent compounds of cuprate superconductors are Mott insulators, and then AFLRO gives rise to a commensurate peak at $[1/2,1/2]$. In Fig. \ref{fig1}, we plot the dynamical spin structure factor $S({\bf k},\omega)$ in the $[k_{x},k_{y}]$ plane at doping $\delta=0.21$ with temperature $T=0.002J$ for energy (a) $\omega =0.1J$, (b) $\omega =0.3J$, and (c) $\omega =0.5J$. When AFLRO is suppressed with doping, two IC magnetic scattering modes separated by a commensurate resonance energy $\omega_{\rm r}\sim 0.3J$ are developed. Well above the magnetic resonance energy $\omega_{\rm r}$, the IC magnetic scattering peaks lie uniformly on a circle of radius of $\bar{\delta}_{\rm IC}'$, and then the distribution of the spectral weight of these IC magnetic scattering peaks is quite isotropic. However, the geometry of the magnetic scattering is energy dependent. In particular, below $\omega_{\rm r}$, although some IC satellite peaks appear along the diagonal direction of BZ, the highest peaks locate at $[(1\pm\bar{\delta}_{\rm IC})/2,1/2]$ and $[1/2,(1\pm\bar{\delta}_{\rm IC})/2]$, and then the main weight of the IC magnetic scattering peaks is in the parallel direction, which leads to an rather anisotropic distribution of the spectral weight of IC magnetic scattering peaks below $\omega_{\rm r}$. To show this energy dependence of the position of the low-energy magnetic scattering peaks clearly, we plot the evolution of the magnetic scattering peaks with energy at $\delta=0.21$ for $T=0.002J$ in Fig. \ref{fig2}, where the hour-glass-shaped dispersion of the low-energy magnetic scattering peaks observed from different families of cuprate superconductors is qualitatively reproduced \cite{Fujita12,Eschrig06,Fong95,Birgeneau89,Cheong91,Yamada98,Dai01,Wakimoto04,Hayden04,Tranquada04,Hinkov04,Bourges05,He01,Bourges00,Arai99,Vignolle07,Stock05,Stock10,Xu09}. In particular, in contrast to the case at energies below $\omega_{\rm r}$, the spin excitations at energies above $\omega_{\rm r}$ disperse almost linearly with energy, in qualitative agreement with the corresponding experimental results \cite{Fujita12,Eschrig06,Fong95,Birgeneau89,Cheong91,Yamada98,Dai01,Wakimoto04,Hayden04,Tranquada04,Hinkov04,Bourges05,He01,Bourges00,Arai99,Vignolle07,Stock05,Stock10,Xu09}.

\subsection{Doping dependence of commensurate resonance}

\begin{figure}[h!]
\includegraphics[scale=0.4]{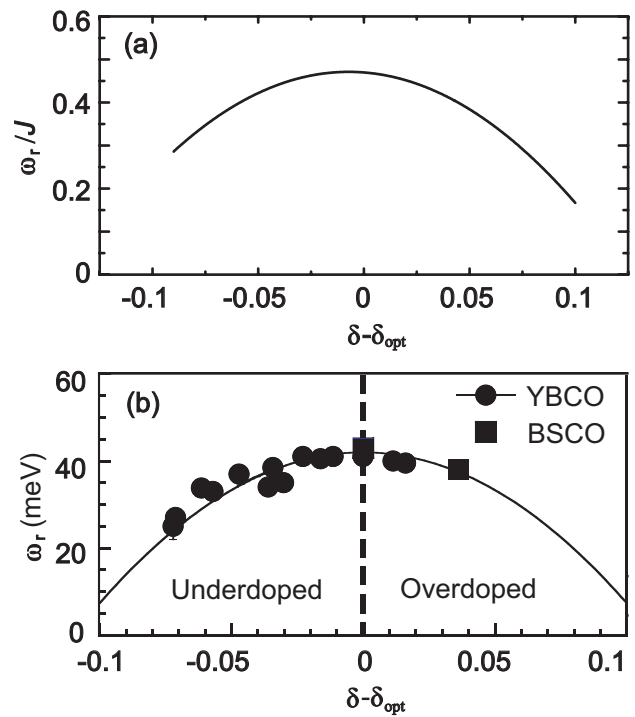}
\caption{(a) The magnetic resonance energy $\omega_{\rm r}$ as a function of doping with $T=0.002J$ for $t/J=2.5$ and $t'/t=0.3$. (b) The experimental results of the doping dependence of the magnetic resonance energy for YBa$_{2}$Cu$_{3}$O$_{6+\delta}$ (circles) and Bi$_{2}$Sr$_{2}$CaCu$_{2}$O$_{8+\delta}$ (squares) taken from Ref. \onlinecite{Bourges05}.\label{fig3}}
\end{figure}

The commensurate resonance energy $\omega_{\rm r}$ in Fig. \ref{fig1}b is strongly doping dependent. For a better understanding of the evolution of $\omega_{\rm r}$ with doping, we have made a series of calculations for $S({\bf k},\omega)$ throughout the entire SC dome, and the result of $\omega_{\rm r}$ as a function of doping with $T=0.002J$ is plotted in Fig. \ref{fig3}a. For comparison, the experimental results \cite{Bourges05} of the doping dependence of the magnetic resonance energy shown in Fig. \ref{fig3}b are obtained from the  cuprate superconductors YBa$_{2}$Cu$_{3}$O$_{6+\delta}$ (circles) and Bi$_{2}$Sr$_{2}$CaCu$_{2}$O$_{8+\delta}$ (squares), respectively. It is shown clearly that in analogy to the domelike shape of the doping dependence of $T_{\rm c}$, the maximal $\omega_{\rm r}$ occurs around the optimal doping, and then decreases in both the underdoped and the overdoped regimes, also in qualitative agreement with the experimental results \cite{Fujita12,Bourges05}. In particular, in the optimal doping $\delta_{\rm opt}=0.15$, the anticipated resonance energy $\omega_{\rm r}=0.46J\approx 46$ meV is not too far from the resonance energy $\omega_{\rm r}\approx 41$ meV observed in the optimally doped YBa$_{2}$Cu$_{3}$O$_{6+\delta}$  \cite{Fujita12,Fong95,Dai01,Bourges05}.

\subsection{Evolution of high-energy spin excitations with doping}

\begin{figure*}[t!]
\includegraphics[scale=0.9]{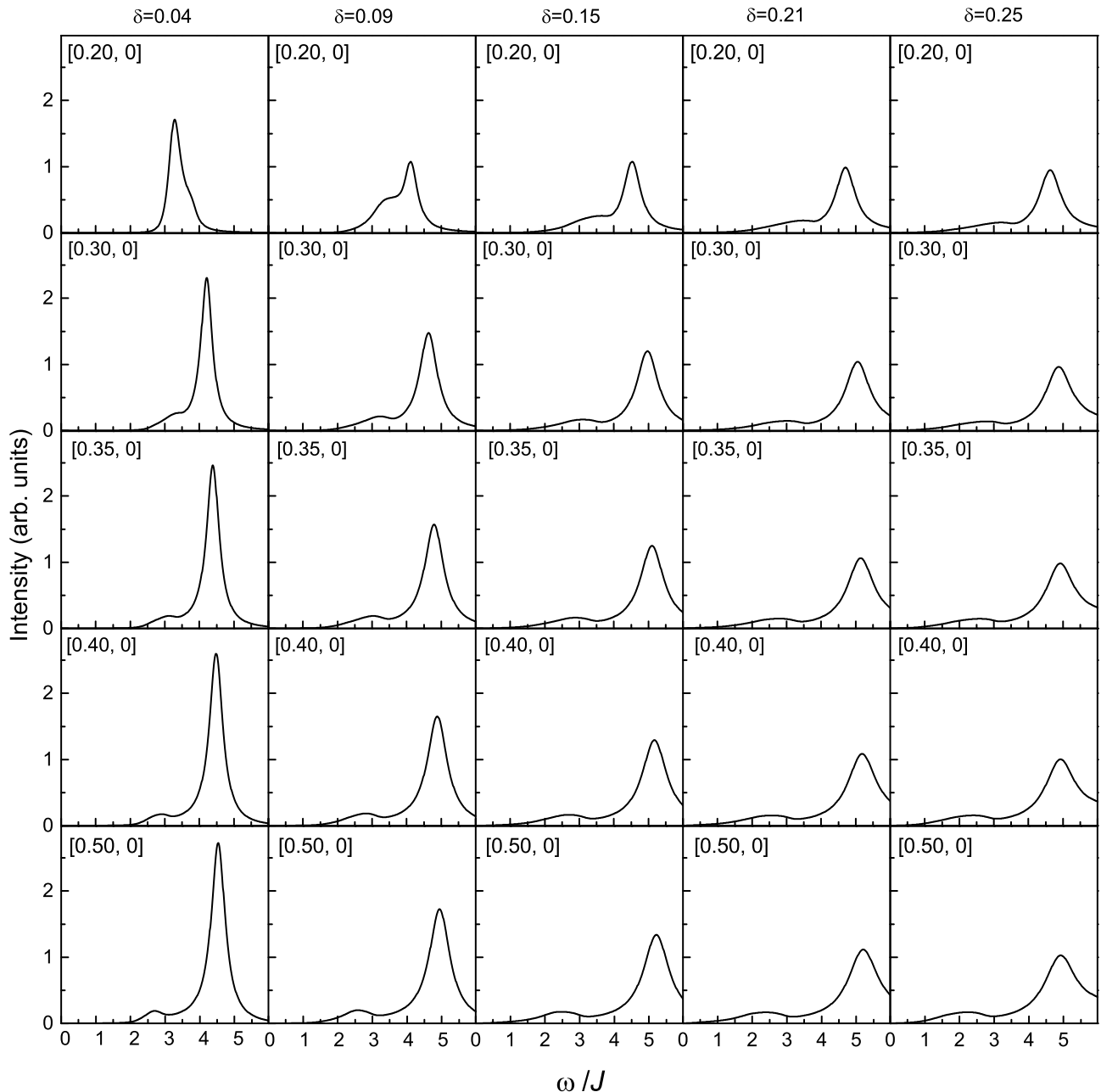}
\caption{The dynamical spin structure factor $S({\bf k},\omega)$ as a function of energy along the ${\bf k}=[0,0]$ to ${\bf k}=[0.5,0]$ direction of the Brillouin zone at $\delta=0.04$, $\delta=0.09$, $\delta=0.15$, $\delta=0.21$, and $\delta=0.25$ with $T=0.002J$ for $t/J=2.5$ and $t'/t=0.3$. \label{fig4}}
\end{figure*}

Now we turn to discuss the striking properties of the high-energy magnetic scattering. For a comparison of the high-energy spin excitations at different doping levels just as it has
been done in the RIXS experiments \cite{Dean14,Tacon11,Dean13,Dean13a,Tacon13}, we plot the dynamical spin structure factor $S({\bf k},\omega)$ as a function of energy along the ${\bf k}=[0,0]$ to ${\bf k}=[0.5,0]$ direction of BZ with $T=0.002J$ at $\delta=0.04$ (non-SC regime), $\delta=0.09$, $\delta=0.15$, $\delta=0.21$, and $\delta=0.25$ in Fig. \ref{fig4}. Obviously, our present theoretical result captures the qualitative feature of the high-energy spin excitations observed experimentally on cuprate superconductors \cite{Dean14,Tacon11,Dean13,Dean13a,Tacon13}. The high-energy spin excitations persist across the whole doping phase diagram with comparable spectral weight and similar energies, i.e., in contrast to the dramatic change of the low-energy spin excitations with doping, the high-energy spin excitations retain roughly constant energy as a function of doping, and the shapes of these high-energy magnetic scattering peaks in the heavy overdoped regime are very similar to those in the lightly doped and underdoped regimes, although the width of the high-energy spin excitations increases continuously with doping, consistent with the spin excitation being damped by the increasing doping. Furthermore, for example, the magnetic scattering peak on energy scale of $4.7J$ appears in the ${\bf k}=[0.2,0]$ point at $\delta=0.21$, while the peak on the energy scale of $5.0J$ emerges in the ${\bf k}=[0.3,0]$ point, reflecting the dispersive nature of the high-energy spin excitations along the ${\bf k}=[0,0]$ to ${\bf k}=[0.5,0]$ direction. In particular, this dispersion relation of the high-energy spin excitations in the overdoped regime resembles those in the lightly doped and underdoped regimes. Our present results also show that within the framework of the kinetic energy driven SC mechanism \cite{Feng0306}, the mediating spin excitations in the SC-state that are coupled to the conducting charge carriers, have energy greater than the charge carrier pair energy.

\subsection{Dispersion of spin excitations}

\begin{figure}[h!]
\includegraphics[scale=0.4]{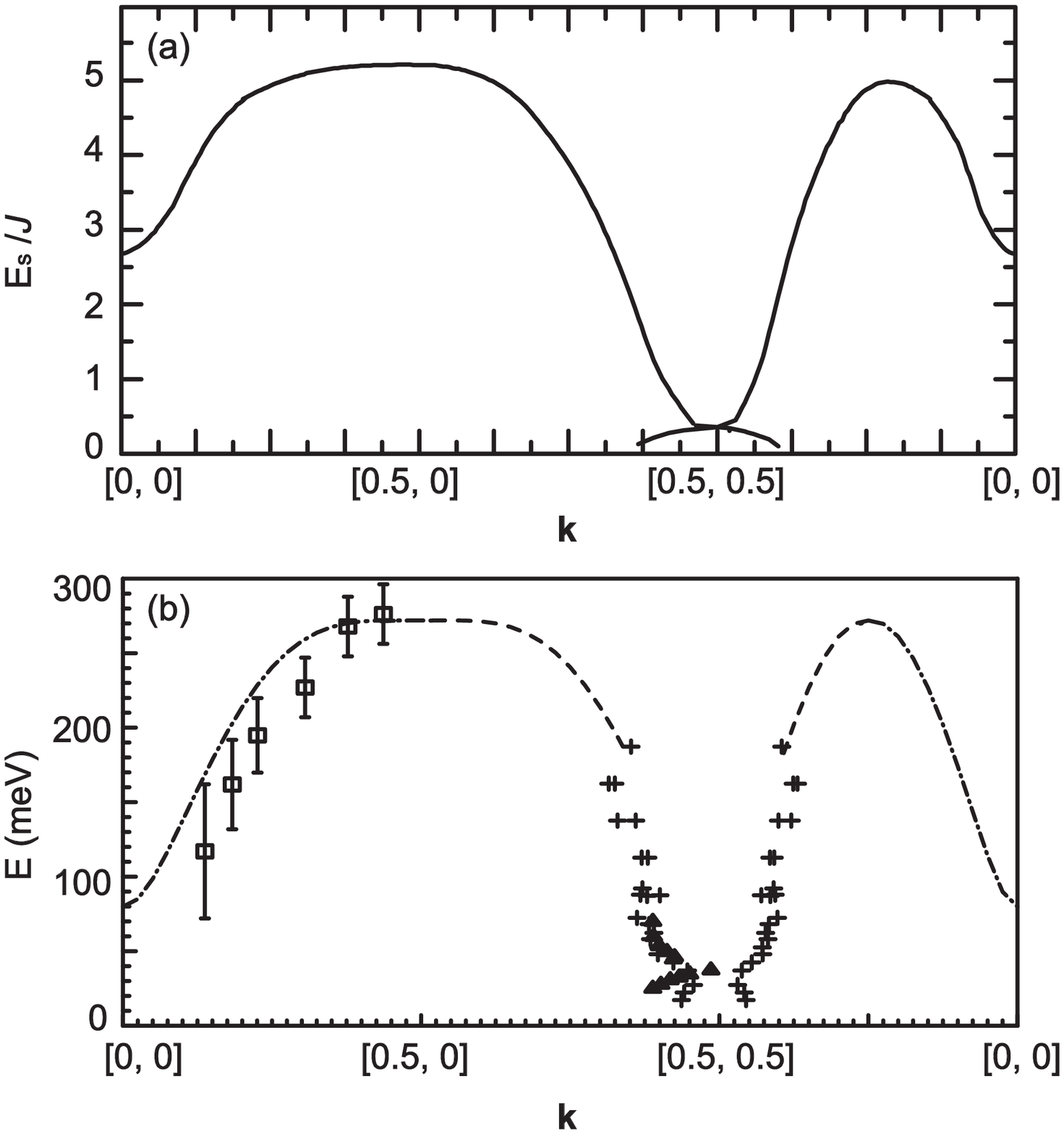}
\caption{(a) The dispersion of the spin excitations along the high symmetry directions of the Brillouin zone at $\delta=0.21$ for $t/J=2.5$ and $t'/J=0.3$ with $T=0.002J$. (b) The experimental result of the dispersion of the spin excitations along the high symmetry directions of the Brillouin zone for YBa$_{2}$Cu$_{3}$O$_{6+\delta}$ taken from Ref. \onlinecite{Vojta11}. \label{fig5}}
\end{figure}

To determine the overall spin excitation spectrum $E_{\rm s}({\bf k})$ in Eq. (\ref{DSSF}), we have performed a {\it self-consistent} calculation,
\begin{eqnarray}
\omega^{2}=\omega^{2}_{\bf k}+B_{{\bf k}}{\rm Re}\Sigma^{({\bf s})}({\bf k},\omega),
\end{eqnarray}
at different momenta in the whole doping phase diagram, and the results show that the spin excitations are well defined at all momenta. In Fig. \ref{fig5}a, we plot $E_{\rm s}({\bf k})$ as a function of momentum along the high symmetry directions of BZ at $\delta=0.21$ with $T=0.002J$. For comparison, the experimental result \cite{Vojta11} of $E_{\rm s}({\bf k})$ along the high symmetry directions of BZ shown in Fig. \ref{fig5}b is obtained from the cuprate superconductor YBa$_{2}$Cu$_{3}$O$_{6+\delta}$. Our present calculations reproduce qualitatively the overall dispersion of the spin excitations in cuprate superconductors \cite{Vojta11}. In comparison with the spin excitation spectrum (the spin wave) of the parent compounds of cuprate superconductors \cite{Fujita12,Hayden91,Hayden96,Coldea01}, it is shown that the spin excitation spectrum in the doped regime has been renormalized due to the presence of the charge carrier-spin interaction in the kinetic energy of the $t$-$J$ model (\ref{cssham}). However, the charge carrier doping does not uniformly renormalizes the dispersion of the spin excitations. In particular, the low-energy magnetic correlations are strongly reorganized, where two IC components of the spin excitation spectrum are separated by the commensurate resonance energy $\omega_{\rm r}$, which therefore leads to an hour-glass-shaped dispersion of the magnetic scattering peaks \cite{Fujita12,Eschrig06,Fong95,Birgeneau89,Cheong91,Yamada98,Dai01,Wakimoto04,Hayden04,Tranquada04,Hinkov04,Bourges05,He01,Bourges00,Arai99,Vignolle07,Stock05,Stock10,Xu09} as shown in Fig. \ref{fig2}. Moreover, the spin excitations at energies well above $\omega_{\rm r}$ disperse almost linearly with energy, which is similar to spin wave with a finite gap, reflecting a fact that the charge carrier doping strongly renormalizes the spin excitations at energies below $\omega_{\rm r}$, but has a modest effect on the spin excitation dispersion at energies above $\omega_{\rm r}$ \cite{Fujita12}. However, in contrast to the case of the low-energy spin excitations, the dispersion of the high-energy spin excitations in cuprate superconductors is strikingly similar to that of their parent compounds \cite{Vojta11,Dean14,Tacon11,Dean13,Dean13a,Tacon13}.

An explanation of unusual magnetic correlations in cuprate superconductors in the SC-state can be found from the spin self-energy $\Sigma^{({\rm s})}({\bf k}, \omega)$ in Eq. (\ref{SSF}) obtained directly from the charge carrier-spin interaction in the kinetic energy of the $t$-$J$ model (\ref{cssham}). This follows from a fact that the dynamical spin structure factor $S({\bf k},\omega)$ in Eq. (\ref{DSSF}) has a well-defined resonance character, where $S({\bf k},\omega)$ exhibits peaks when the incoming neutron energy $\omega$ is equal to the spin excitation energy $E_{\rm s}({\bf k})$, i.e.,
\begin{eqnarray}
\omega^{2}-\omega_{{\bf k}_{\rm c}}^{2}-B_{{\bf k}_{\rm c}}{\rm Re}\Sigma^{({\rm s})}({\bf k}_{\rm c},\omega)=\omega^{2}-E^{2}_{\rm s}({\bf k}_{\rm c})\sim 0,
\end{eqnarray}
for certain critical wave vectors ${\bf k}_{\bf c}$, the magnetic scattering peaks appear, and then the weights of these peaks are dominated by the inverse of the imaginary part of the spin self-energy $1/{\rm Im}\Sigma^{({\rm s})}({\bf k}_{\rm c},\omega)$. In other words, the positions of the magnetic scattering peaks are determined by both the spin excitations energy $E_{\rm s}({\bf k})$ and the imaginary part of the spin self-energy ${\rm Im} \Sigma^{({\rm s})}({\bf k}_{\rm c},\omega)$. At half-filling, the low-energy magnetic scattering peak locates at the AF wave vector $[1/2,1/2]$, so the commensurate AF peak appears there. However, away from half-filling, the doped charge carriers disturb the AF background. In particular, within the framework of the kinetic energy driven SC mechanism, as a result of the self-consistent interplay between the charge carriers and spins, unusual magnetic correlations are developed. As mentioned above, the spin self-energy $\Sigma^{({\rm s})}({\bf k},\omega)$ in Eq. (\ref{SSF}) is obtained in terms of the full charge carrier diagonal Green's function (\ref{BCSHDGF}) and off-diagonal Green's function (\ref{BCSHODGF}), and renormalizes the spin excitations. However, in the charge carrier quasiparticle spectrum $E_{{\rm h}{\bf k}}=\sqrt{\bar{\xi}^{2}_{{\bf k}}+\mid\bar{\Delta}_{\rm hZ} ({\bf k}) \mid^{2}}$ in the full charge carrier diagonal Green's function (\ref{BCSHDGF}) and off-diagonal Green's function (\ref{BCSHODGF}), the maximal $|\bar{\xi}_{{\bf k}}|$ appears around the nodal region, and then $-2J <\bar{\xi}_{{\bf k}}< 2J$ occurs at the end of the SC dome \cite{Feng0306,Guo06,Feng08}, i.e., $\bar{\xi}_{{\bf k}}$ has an effective band width $W_{\rm h}\sim 2J$ around the charge carrier Fermi level at the end of the SC dome, and then decreases with decreasing doping, while the d-wave charge carrier pair gap $\bar{\Delta}_{\rm h}({\bf k})$ vanishes on the gap nodes, and the d-wave charge carrier pair gap parameter $\bar{\Delta}_{\rm h}$ has a domelike shape of the doping dependence with the maximal $\bar{\Delta}_{\rm h}\sim 0.2J$ occurred around the optimal doping \cite{Feng0306,Guo06,Feng08}. These properties of $\bar{\xi}_{{\bf k}}$ and $\bar{\Delta}_{\rm h}({\bf k})$ lead to that the spin self-energy in (\ref{SSF}) strongly renormalizes the spin excitations at energies below $W_{\rm h}$, but has a weak effect on the spin excitations at energies above $W_{\rm h}$. This is why the magnetic correlations at energies below $W_{\rm h}$ are strongly reorganized, while the high-energy spin fluctuations bear a striking resemblance to those found in the parent compound. Furthermore, as seen from the spin self-energy (\ref{SSF}), there are two parts of the charge carrier quasiparticle contribution to the spin self-energy renormalization. The contribution from the first term of the right-hand side in Eq. (\ref{SSF}) mainly comes from the mobile charge carrier quasiparticles, and the coherence factor for this process is given in Eq. (\ref{CFS1}). This process mainly leads to low-energy IC magnetic scattering, and can persist into the normal-state. However, the additional contribution from the second term of the right-hand side in Eq. (\ref{SSF}) originates from the creation of charge carrier pairs, and the coherence factor for this additional process is given in Eq. (\ref{CFS2}). This additional process occurs in the SC-state {\it only}, and gives a dominant contribution to the commensurate resonance \cite{Fong95}. This is why the commensurate resonance is intimately related to superconductivity, and then appears in the SC-state {\it only}. Within the kinetic energy driven SC mechanism  \cite{Feng0306,Guo06,Feng08}, superconductivity results when charge carriers pair up into charge carrier pairs, while the charge carrier pair gap parameter $\bar{\Delta}_{\rm h}$ and $T_{\rm c}$ have domelike shape of the doping dependence, which leads to that the resonance energy $\omega_{\rm r}$ shows the same domelike shape of the doping dependence as $\bar{\Delta}_{\rm h}$ and $T_{\rm c}$.

\section{Quantitative characteristics in the normal-state}\label{normal-state}

We now address the dynamical spin response of cuprate superconductors in the normal-state. At the temperature $T>T_{\rm c}$, the charge carrier pair gap parameter $\bar{\Delta}_{\rm h}=0$, then superconductivity disappears, and the system becomes a strange metal. In this case, the SC-state dynamical spin structure factor $S({\bf k},\omega)$ in Eq. (\ref{DSSF}) is reduced to that in the normal-state, where the spin self-energy is obtained in terms of the collective charge carrier mode in the particle-hole channel {\it only}, and can be obtained explicitly as \cite{Feng98,Feng04},
\begin{eqnarray}\label{SSFN}
\Sigma^{({\rm s})}({\bf k},\omega)&=&-{2\over N^{2}}\sum_{\bf pq}\Omega({\bf k},{\bf p},{\bf q})\nonumber\\
&\times&{F^{\rm (n)}({\bf k},{\bf p},{\bf q})\over\omega^{2}-[\omega_{{\bf q}+{\bf k}}- (\bar{\xi}_{{\bf p}+{\bf q}}-\bar{\xi}_{{\bf p}})]^{2}},
\end{eqnarray}
with the function,
\begin{eqnarray}
F^{\rm (n)}({\bf k},{\bf p},{\bf q})&=&[\omega_{{\bf q}+{\bf k}}-(\bar{\xi}_{{\bf p}+{\bf q}}-\bar{\xi}_{{\bf p}})]\nonumber\\
&\times&\{n_{\rm B}(\omega_{{\bf q}+{\bf k}})[n_{\rm F}(\bar{\xi}_{{\bf p}})-n_{\rm F}(\bar{\xi}_{{\bf p}+{\bf q}})]\nonumber\\
&-&[1-n_{\rm F}(\bar{\xi}_{{\bf p}})]n_{\rm F}(\bar{\xi}_{{\bf p}+{\bf q}})\}.
\end{eqnarray}

\subsection{Low-energy incommensurate spin fluctuation}

\begin{figure}[h!]
\includegraphics[scale=0.85]{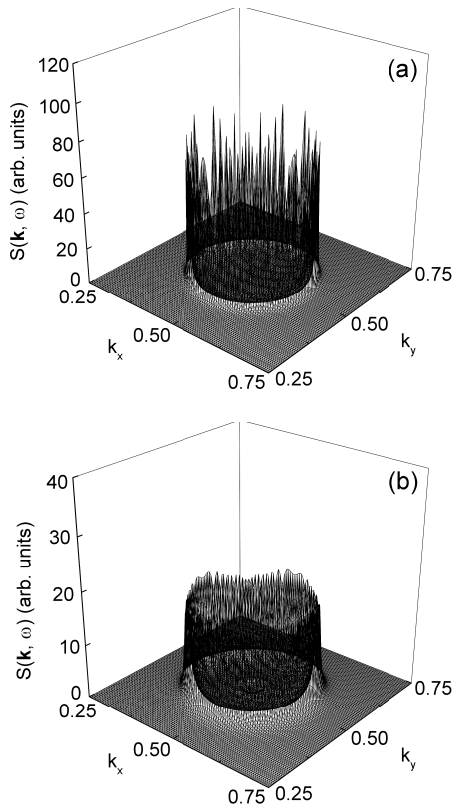}
\caption{The normal-state dynamical spin structure factor $S({\bf k},\omega)$ in the $[k_{x},k_{y}]$ plane at $\delta=0.21$ with $T=0.09J$ in (a) $\omega=0.1J$ and (b) $\omega=0.3J$ for $t/J=2.5$ and $t'/t=0.3$. \label{fig6a}}
\end{figure}

\begin{figure*}[t!]
\includegraphics[scale=0.9]{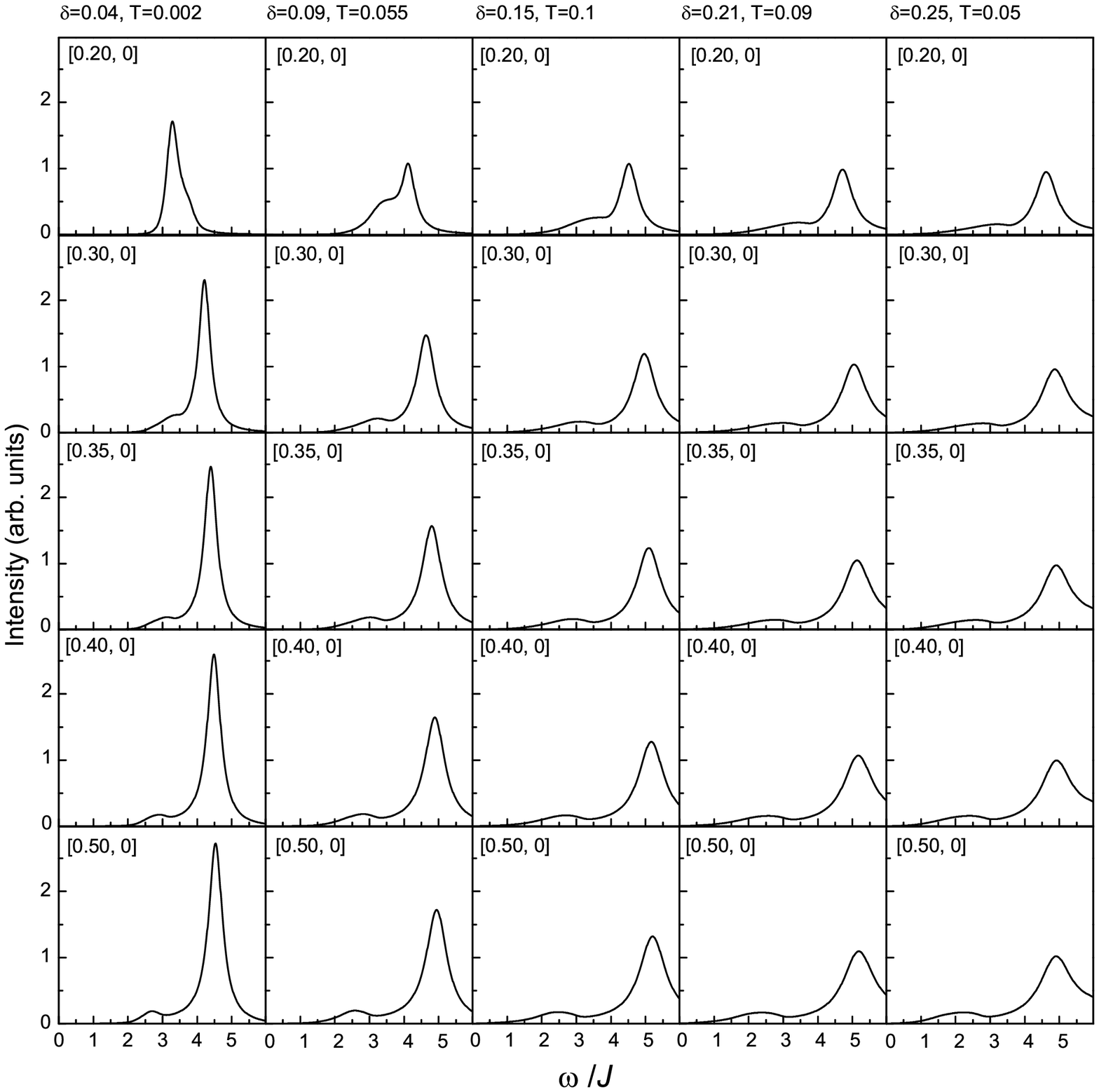}
\caption{The normal-state dynamical spin structure factor $S({\bf k},\omega)$ as a function of energy along the ${\bf k}=[0,0]$ to ${\bf k}=[0.5,0]$ direction at $\delta=0.04$ with $T=0.002J$, $\delta=0.09$ with $T=0.05J$, $\delta=0.15$ with $T=0.1J$, $\delta=0.21$ with $T=0.09J$, and $\delta=0.25$ with $T=0.05J$ for $t/J=2.5$ and $t'/t=0.3$. \label{fig7}}
\end{figure*}

Within the framework of the kinetic energy driven SC mechanism, the calculated \cite{Huang13} $T_{\rm c}\sim 0.065J$ at the doping concentration $\delta=0.21$. To show how the spin excitations evolve with temperature from the SC-state in Fig. \ref{fig1} to the normal-state, we plot the normal-state $S({\bf k},\omega)$ in the $[k_{x},k_{y}]$ plane at $\delta=0.21$ with $T=0.09J$ in (a) $\omega=0.1J$ and (b) $\omega=0.3J$ in Fig. \ref{fig6a}. Comparing it with Fig. \ref{fig1} for the same set of parameters except for $T=0.09J$, we see that the commensurate resonance is absent from the present normal-state, reflecting a fact that that only the low-energy IC spin fluctuations in the SC-state can persist into the normal-state, while the commensurate resonance is a new feature that appears in the SC-state only \cite{Fujita12,Fong95,Dai01,Bourges05}. Moreover, the low-energy IC magnetic scattering peaks lie on a circle of radius $\bar{\delta}_{\rm IC}$. Although some IC satellite peaks along the diagonal direction appear, the main weight of the IC magnetic scattering peaks is in the parallel direction as in the case of the SC-state, and these peaks along the parallel direction are located at $[(1\pm\bar{\delta}_{\rm IC})/2,1/2]$ and $[1/2,(1\pm\bar{\delta}_{\rm IC})/2]$. The IC magnetic scattering peaks are very sharp at lower energies, however, they broaden and weaken in amplitude as the energy increase. This reflects a fact that the width of the spin excitations in the normal-state increases with increasing energies, in other words, the lifetime of the spin excitations decreases with increasing energies. In particular, the dynamical spin structure factor spectrum has been used to extract the doping dependence of the incommensurability parameter $\bar{\delta}_{\rm IC}$, and the results \cite{Feng98,Feng04} show clearly that $\bar{\delta}_{\rm IC}$ increases progressively with doping at the lower doped regime, but saturates at the higher doped regime, in qualitative agreement with experiments \cite{Yamada98,Enoki13}.

\subsection{High-energy magnetic scattering}

To analyze the evolution of the high-energy spin excitations with doping in the normal-state, we have performed a calculation of the normal-state $S({\bf k},\omega)$ at the different doping levels with the temperature well above the corresponding $T_{\rm c}$, and the results of the normal-state $S({\bf k},\omega)$ as a function of energy along the ${\bf k}=[0,0]$ to ${\bf k}= [0.5,0]$ direction of BZ at $\delta=0.04$ with $T=0.002J$, $\delta=0.09$ with $T=0.05J$, $\delta=0.15$ with $T=0.1J$, $\delta=0.21$ with $T=0.09J$, and $\delta=0.25$ with $T=0.05J$ are plotted in Fig. \ref{fig7}. In comparison with the corresponding results of the SC-state in Fig. \ref{fig4}, we show the existence of the high-energy spin excitations in the normal-state for all doping levels. Although the high-energy spin excitations retain roughly constant energy as a function of doping, the width of these high-energy spin excitations increases with increasing doping. In particular, these high-energy spin excitations in the normal-state, in their overall dispersion, their spectral weight, and the shapes of the magnetic scattering peaks, are striking similar to those in the corresponding SC-state, although the magnetic scattering peak in the normal-state is softening and broadening. These results are also in qualitative agreement with the experimental results \cite{Dean14,Tacon11,Dean13,Dean13a,Tacon13}.

\subsection{Overall dispersion of spin excitations}

\begin{figure}[h!]
\includegraphics[scale=0.43]{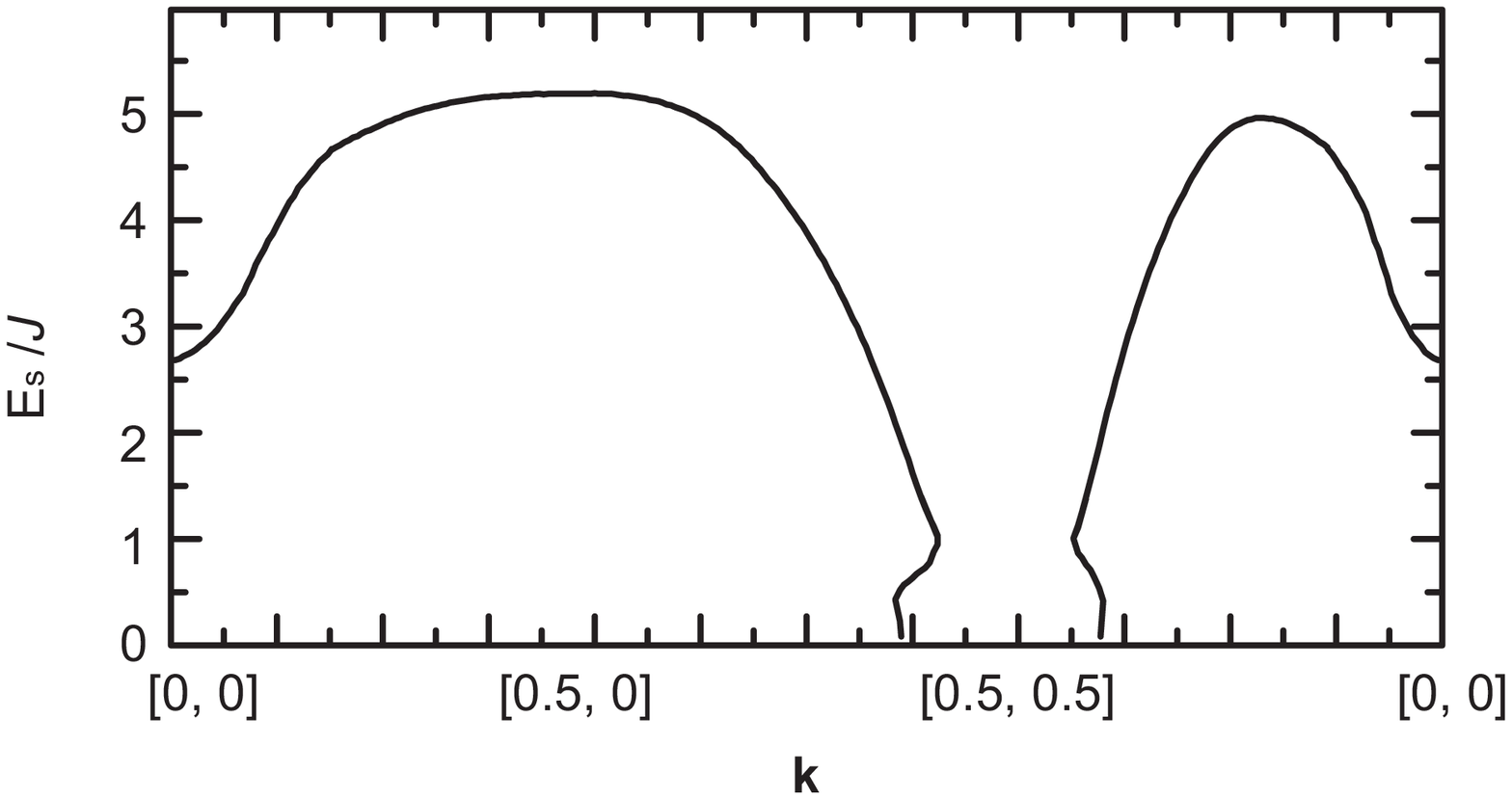}
\caption{The dispersion of the spin excitations in the normal-state along the high symmetry directions of the the Brillouin zone at $\delta=0.21$ for $t/J=2.5$ and $t'/J=0.3$ with $T=0.09J$. \label{fig8}}
\end{figure}

For a complement of the analysis of the nature of the spin excitations in the normal-state, we plot the spin excitation spectrum $E_{\rm s}({\bf k})$ in the normal-state as a function of momentum along the high symmetry directions with $T=0.09J$ at $\delta=0.21$ in Fig. \ref{fig8}. In comparison with the corresponding results of the spin excitation spectrum in Fig. \ref{fig5}a in the SC-state, it is shown clearly that the dispersion relation of the high-energy spin excitations in the normal-state resemble those in the SC-state. On the other hand, although the low-energy spin excitations in the normal-state exhibit a dispersion quite similar to the hour-glass behavior as in the SC-state, the positions of the magnetic scattering peaks at the waist of this {\it hour glass} drift away from $[1/2,1/2]$, and then the IC-shaped dispersion emerges in the whole low-energy range.

The essential physics of the dynamical spin response of cuprate superconductors in the normal-state is the same as in the SC-state, and also can be attributed to the spin self-energy effects which arise directly from the charge carrier-spin interaction in the kinetic energy of the $t$-$J$ model (\ref{cssham}). However, in the SC-state, the spin self-energy renormalization in Eq. (\ref{SSF1}) is due to the charge carrier bubbles in both the charge carrier particle-hole and particle-particle channels as mentioned in Sec. \ref{framework}, where both processes (\ref{CFS}) from the mobile charge carrier quasiparticles and the creation of charge carrier pairs contribute to the spin self-energy renormalization. This is different from the case in the normal-state, where the spin self-energy renormalization in Eq. (\ref{SSFN}) is due to the charge carrier bubble in the charge carrier particle-hole channel only, i.e., only process from the mobile charge carrier quasiparticles contributes to the spin self-energy renormalization. This difference leads to an absence of the commensurate resonance from the normal-state. These results in the normal-state confirm again that the commensurate resonance appears in the SC-state only, while the low-energy IC magnetic scattering can persist into the normal-state. In particular, the effective band width $W_{\rm h}$ of $\bar{\xi}_{{\bf k}}$ in the normal-state is almost the same as that in the corresponding SC-state, and then in analogy to the case in the SC-state, the spin self-energy in (\ref{SSFN}) strongly renormalizes the spin excitations at energies below $W_{\rm h}$, but has a weak effect on the spin excitations at energies above $W_{\rm h}$. This is why the high-energy spin excitations in the normal-state retain roughly constant energy as a function of doping, with the shape of the magnetic scattering peaks, spectral weights and dispersion relations comparable to those in the corresponding SC-state. Since the height of the magnetic scattering peaks is determined by damping, it is fully understandable that they are suppressed as the energy and temperature are increased.

\section{Summary and discussion}\label{conclusions}

\begin{figure}[h!]
\includegraphics[scale=0.4]{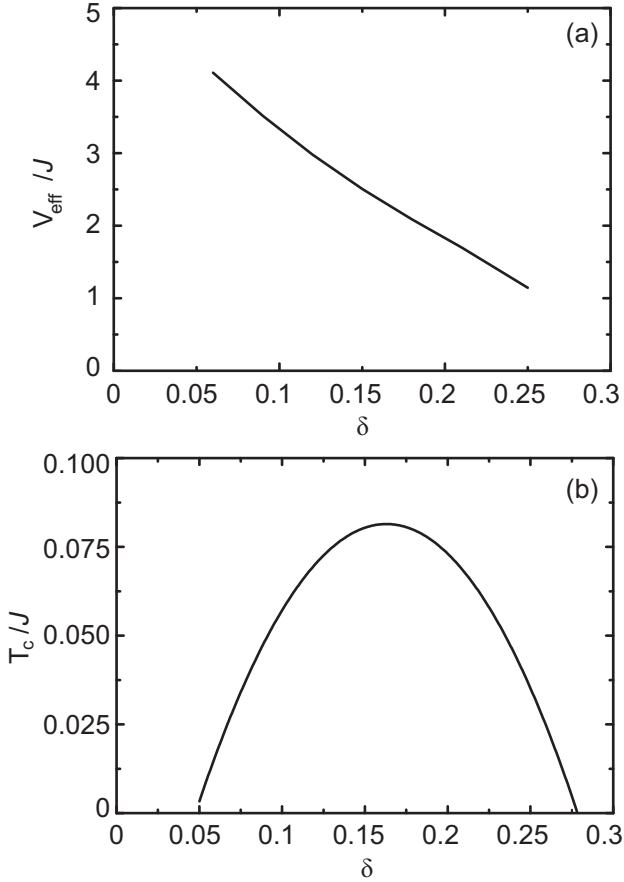}
\caption{(a) The coupling strength at $T=0.002J$ and (b) the superconducting transition temperature as a function of doping for $t/J=2.5$ and $t'/t=0.3$. \label{fig9ab}}
\end{figure}

In cuprate superconductors, a small density of charge carriers is sufficient to destroy AFLRO, however, the doped charge carriers and magnetic exchange-coupled spins organize themselves in a cooperative way to enhance both charge carrier mobility and AFSRO correlations, and then the spin excitations in the spin liquid state with AFSRO appear to survive up to the high-energy \cite{Fujita12,Dean14}. In the doped regime without AFLRO, we \cite{Feng0306} have shown within the framework of the kinetic energy driven SC mechanism that the charge carrier-spin interaction directly from the kinetic energy of the $t$-$J$ model (\ref{cssham}) by exchanging spin excitations in higher powers of the doping concentration induces the SC-state. In particular, although the kinetic energy increases with increasing doping, the strength $V_{\rm eff}$ of the charge carrier attractive interaction mediated by spin excitations in the kinetic energy driven SC mechanism smoothly decreases upon increasing doping from a strong-coupling case in the underdoped regime to a weak-coupling side in the overdoped regime as shown in Fig. \ref{fig9ab}a \cite{Huang13,Kordyuk10}, reflecting a decrease of the intensity of the spin excitations with increasing doping. In the underdoped regime, the coupling strength $V_{\rm eff}$ is very strong, and then most charge carriers can be bound to form charge carrier pairs. In this case, the number of the charge carrier pairs increases with increasing doping, which leads to an increase of $T_{\rm c}$ with increasing doping. However, in the overdoped regime, the coupling strength $V_{\rm eff}$ is relatively weak. In this case, not all charge carriers can be bound to form charge carrier pairs by the weakly attractive interaction, and therefore the number of the charge carrier pairs decreases with increasing doping, this leads to a decrease of $T_{\rm c}$ with increasing doping. In other words, the reduction in $T_{\rm c}$ on the overdoped side is driven by a reduction in the coupling strength $V_{\rm eff}$ of the pairing interaction. In particular, the optimal doping is a balance point, where the number of charge carrier pairs and coupling strength are optimally matched. This is why $T_{\rm c}$ takes a domelike shape with the underdoped and overdoped regimes on each side of the optimal doping, where $T_{\rm c}$ reaches its maximum as shown in Fig. \ref{fig9ab}b \cite{Huang13}. On the other hand, our present results show that the evolution of the spin excitations with doping in cuprate superconductors can be well described within the framework of the kinetic energy driven SC mechanism from low-energy to high-energy. However, since the {\it highest}-energy spin excitations in the SC-state retain roughly constant energy as a function of doping, with spectral weights and dispersion relations almost same with those in the corresponding normal-state, it is shown clearly from our present results based on the kinetic energy driven SC mechanism together with the data  measured by the RIXS experiments \cite{Dean14,Tacon13} that the {\it highest}-energy spin excitations are unlikely to be a major factor in the pairing interaction. In this case, the coupling strength of the pairing interaction comes mostly from the {\it lower}-energy spin excitations, which change with doping and persist across the SC dome while retaining sufficient intensity. In particular, as a natural consequence of the suppression of the intensity of the magnetic scattering peaks with increasing doping, the coupling strength $V_{\rm eff}$ decreases with increasing doping, and in parallel with that $T_{\rm c}$ reduces in the overdoped side.

To summarize, within the framework of the kinetic energy driven SC mechanism, we have studied the dynamical spin response of cuprate superconductors from low-energy to high-energy. The spin self-energy in the SC-state is evaluated explicitly in terms of the collective charge carrier modes in the particle-hole and particle-particle channels, and then employed to calculate the dynamical spin structure factor. Our results show the existence of damped but well-defined dispersive spin excitations in the whole doping phase diagram. In particular, the low-energy spin excitations in the SC-state have an hour-glass-shaped dispersion, with commensurate resonance that originates from the process of the creation of charge carrier pairs, and appears in the SC-state only, while the low-energy IC spin fluctuations are dominated by the process from the mobile charge carrier quasiparticles, and therefore can persist into the normal-state. The high-energy spin excitations in the SC-state on the other hand retain roughly constant energy as a function of doping, with spectral weights and dispersion relations comparable to those in the corresponding normal-state, although the magnetic scattering peak in the normal-state is softening and broadening. Our theory also shows that unusual magnetic correlations in cuprate superconductors are ascribed purely to the spin self-energy effects which arise directly from the charge carrier-spin interaction in the kinetic energy of the system.

\acknowledgments

The authors would like to thank Dr. Zheyu Huang and Dr. Huaisong Zhao for helpful discussions. LK and SF are supported by the funds from the Ministry of Science and Technology of China under Grant Nos. 2011CB921700 and 2012CB821403, and the National Natural Science Foundation of China under Grant No. 11274044, and YL is supported by the Science Foundation of Hengyang Normal University under Grant No. 13B44.

\end{document}